\documentclass[aps,prb,a4paper,10pt,twocolumn,showpacs,floatfix,superscriptaddress,preprintnumbers,longbibliography]{revtex4-2}
\usepackage{float}
\usepackage{dcolumn,graphicx,color,booktabs,microtype,afterpage}
\usepackage{amssymb}
\usepackage{amsmath}
\usepackage[english]{babel}
\usepackage[charter,greekuppercase=italicized]{mathdesign}
\usepackage{sidecap}
\usepackage[mathlines]{lineno}

\graphicspath{{./}{figure/}}
\renewcommand{\tablename}{Table}
\makeatletter\renewcommand{\fnum@figure}[1]{\figurename~\thefigure.~}\makeatother
\makeatletter\renewcommand{\fnum@table}[1]{\tablename~\thetable.}\makeatother

\newcount\hh \newcount\mm
\hh=\time \divide\hh by 60
\mm=\hh \multiply\mm by 60 \mm=-\mm
\advance\mm by \time
\def\now{\number\hh:\ifnum\mm<10{}0\fi\number\mm}

\usepackage[colorlinks,plainpages=false,linkcolor=blue,urlcolor=blue,citecolor=blue,pdfpagemode=UseNone,pdfstartview=FitBH]{hyperref}
\usepackage{nicefrac}

\newcommand{\tcr}[1]{\textcolor{black}{#1}}

\begin{document}

\makeatletter\renewcommand{\ps@plain}{%
\def\@evenhead{\hfill\itshape\rightmark}%
\def\@oddhead{\itshape\leftmark\hfill}%
\renewcommand{\@evenfoot}{\hfill\small{--~\thepage~--}\hfill}%
\renewcommand{\@oddfoot}{\hfill\small{--~\thepage~--}\hfill}%
}\makeatother\pagestyle{plain}

\preprint{\textit{Preprint: \today, \now}} 

\title{Anomalous Hall effect and rich magnetic phase diagram of Mn$_{100-x}$Rh$_x$ epitaxial films
}

%
%

\author{Cong Wang}
\affiliation{Key Laboratory of Polar Materials and Devices (MOE), School of Physics and Electronic Science, East China Normal University, Shanghai 200241, China}
\author{Zheng Li}
\affiliation{Key Laboratory of Polar Materials and Devices (MOE), School of Physics and Electronic Science, East China Normal University, Shanghai 200241, China} 
%
%
%
%
\author{Jing Meng}
\affiliation{Key Laboratory of Polar Materials and Devices (MOE), School of Physics and Electronic Science, East China Normal University, Shanghai 200241, China}
\author{Hui Zhang}
\affiliation{Key Laboratory of Polar Materials and Devices (MOE), School of Physics and Electronic Science, East China Normal University, Shanghai 200241, China}
\author{Haoyu Lin}
\affiliation{Key Laboratory of Polar Materials and Devices (MOE), School of Physics and Electronic Science, East China Normal University, Shanghai 200241, China}
\author{Jiyuan Li}
\affiliation{Key Laboratory of Polar Materials and Devices (MOE), School of Physics and Electronic Science, East China Normal University, Shanghai 200241, China}
\author{Kun Zheng}
\affiliation{Key Laboratory of Polar Materials and Devices (MOE), School of Physics and Electronic Science, East China Normal University, Shanghai 200241, China}
%
%
%
%
\author{Yang Xu}
\affiliation{Key Laboratory of Polar Materials and Devices (MOE), School of Physics and Electronic Science, East China Normal University, Shanghai 200241, China}
\author{Tian Shang}\email[Corresponding author:\\]{tshang@phy.ecnu.edu.cn}
\affiliation{Key Laboratory of Polar Materials and Devices (MOE), School of Physics and Electronic Science, East China Normal University, Shanghai 200241, China}
\author{Qingfeng Zhan}\email[Corresponding author:\\]{qfzhan@phy.ecnu.edu.cn}
\affiliation{Key Laboratory of Polar Materials and Devices (MOE), School of Physics and Electronic Science, East China Normal University, Shanghai 200241, China}

%
%
%

%
%
%
\begin{abstract}
A series of Mn$_{100-x}$Rh$_x$ ($20 \le x \le 50$) thin films were epitaxially grown on the MgO substrate using magnetron sputtering technique, and were systematically investigated by magnetization, longitudinal electrical resistivity, and transverse Hall resistivity. After optimizing the growth conditions, phase-pure Mn$_{100-x}$Rh$_x$ films with a cubic CsCl-type structure were obtained, and their magnetic phase diagram was built.  
The manipulation of Rh content leads to a rich magnetic phase diagram, where three different regimes can be identified: 
for $x < 40$, Mn$_{100-x}$Rh$_x$ films undergo a ferromagnetic (FM) transition below $T_\mathrm{C} \approx$ 330-350\,K; for $40 \le x \le 45$, in addition to the FM transition at $T_\mathrm{C} \approx$ 200\,K, Mn$_{100-x}$Rh$_x$ films undergo a FM-to-antiferromagnetic (AFM) transition at $T_\mathrm{N} \approx$ 120\,K; finally for $x > 45$, only one AFM transition at $T_\mathrm{N} \approx$ 150\,K can be tracked. 
All the Mn$_{100-x}$Rh$_x$ films exhibit distinct anomalous Hall effect in their magnetically ordered state, which is most likely due to the intrinsic Berry-curvature mechanism. In addition, all the anomalous Hall transport properties, including the resistivity, conductivity, and angle exhibit a strong correlation with the magnetic properties of Mn$_{100-x}$Rh$_x$ films, which become most evident for $x$ = 35. 
Our systematic investigations suggest a strong correlation between magnetic properties and electronic band topology in Mn$_{100-x}$Rh$_x$ films,
highlighting their great potential for AFM spintronics. 

\end{abstract}

\maketitle\enlargethispage{3pt}

\vspace{-5pt}
\section{\label{sec:Introduction}Introduction}\enlargethispage{8pt}
Antiferromagnetic (AFM) materials with a noncollinear spin structure have attracted great interest due to their exotic physical properties~\cite{smejkal_2022,yan_2020,guo_2025}. One of the representative classes is the Mn$_3$$X$ ($X$ = Ir, Pt, Rh, Ge, Sn) family, where Mn atoms form a kagome lattice with a 120$^\circ$ spin configuration. Despite weak net magnetization, these materials exhibit distinct anomalous Hall effect (AHE) and spin Hall effect (SHE) even at room temperature, mostly arising from the Berry curvatures of the nontrivial electronic bands~\cite{nayak_90,zhang_92,nakatsuji2015_1,xu2025_2,yang2017_5}. 
In particular, Mn$_3$Ir and Mn$_3$Pt have been extensively studied due to their strong spin-orbit coupling and symmetry-governed spin-current anisotropy~\cite{arpaci2021_7,qi2024_8,holanda2020_9, liang2023_10,cao2023_11,qin2023_12}.



Collinear AFM materials such as Mn$_{50}$Pt$_{50}$ and Mn$_{50}$Ir$_{50}$ have been widely applied in the spintronic devices due to their high N\'eel temperature and strong magnetocrystalline anisotropy~\cite{pal_1968,frangou_enhanced_2016}. Their AFM ordering is highly sensitive to the film thickness, chemical disorder, and epitaxial strain~\cite{kangMnPt_2023,liuMnPt_2016,zhou_magnetic_2020}, and they typically serve as an AFM pinning layer in the spin-valve heterostructures~\cite{hase_weak_2001,anderson_2000}. Moreover, recent studies reveal that the spin Hall angle and exchange bias field can be effectively tuned via chemical composition and epitaxial orientation in Mn$_{100-x}$Ir$_x$ films~\cite{zhang_91}.


While Mn-Pt and Mn-Ir binary alloy thin films have been frequently studied, Mn-Rh family remains largely unexplored.  
Bulk Mn$_{50}$Rh$_{50}$ adopts a cubic CsCl-type structure and undergoes a paramagnetic (PM) to AFM transition at $T_\mathrm{N}$ $\approx$ 170 K~\cite{kouvel1963_17}. Similar to the Fe$_{50}$Rh$_{50}$~\cite{zverev_peculiarities_2021}, bulk Mn$_{50}$Rh$_{50}$ alloy exhibits a giant temperature hysteresis ($\Delta$$T$ $\approx$ 130\,K) at $T_\mathrm{N}$ in both temperature-dependent magnetic susceptibility and electrical resistivity~\cite{kouvel1963_17}. Such a cubic phase of Mn$_{100-x}$Rh$_x$ can be stabilized by varying the Rh content from 35 to 50\%~\cite{kainzbauer2020_18}. In addition, a martensitic-like cubic-to-tetragonal distortion accompanied by a $\sim$2\% volume contraction has been found in Mn$_{100-x}$Rh$_x$ (30 $\lesssim x \lesssim$ 50) above room temperature~\cite{kouvel1963_17,kainzbauer2020_18}.
These rich magnetic and structural phase transitions of Mn$_{100-x}$Rh$_x$ alloys resembles the isostructural Fe$_{50}$Rh$_{50}$~\cite{zarkevich_ferh_2018},   
both represent one of the ideal candidate materials for spintronic applications, such as memory resistor~\cite{Marti_2014}, heat-assisted magnetic  recording~\cite{Thiele_2004}, and magnetic refrigeration~\cite{QIAO_2020}. 

In the case of Mn$_{100-x}$Rh$_{x}$ thin films, the studies are limited, which report inconsistent magnetic properties. Polycrystalline Mn$_{50}$Rh$_{50}$ thin film has been found to undergo the magnetic phase transition between AFM and ferromagnetic (FM) states at a temperature around 175\,K and 310\,K during the cooling and heating process, respectively~\cite{chaturvedi2016_20,chaturvedi201620}. However, the epitaxial Mn$_{50}$Rh$_{50}$ film undergoes an AFM transition at $T_\mathrm{N} \approx$ 150\,K, and exhibits a remarkable 360\% enhancement of damping-like spin–orbit torque near the PM-to-AFM transition~\cite{li2025_22}. In addition, the coexistence of AFM and FM orders has been found in Mn$_{50-x}$Fe$_x$Rh$_{50}$ films~\cite{horky2022_21,li202522}.
Such inconsistent results indicate that the magnetic properties of Mn$_{100-x}$Rh$_{x}$ thin films are sensitive to the Mn- or Rh-content, epitaxial strain, and crystallinity. A systematic investigation on Mn$_{100-x}$Rh$_x$ films with different compositions is highly desirable to reveal their intrinsic magnetic properties.

Here, a series of Mn$_{100-x}$Rh$_x$ ($20 \le x \le 50$) epitaxial thin films were grown by varying the Mn or Rh contents, and we report a systematic study of their magnetic and transport properties by means of magnetization, electrical resistivity, and Hall resistivity measurements. A rich magnetic phase diagram was constructed for Mn$_{100-x}$Rh$_x$ films, which exhibit distinct AHE in the magnetically ordered states. We propose that the observed AHE 
in Mn$_{100-x}$Rh$_x$ films is most likely attributed to the Berry curvatures in the momentum space.

\begin{figure}[!thp]
	\centering
	\includegraphics[width=0.48\textwidth,angle=0]{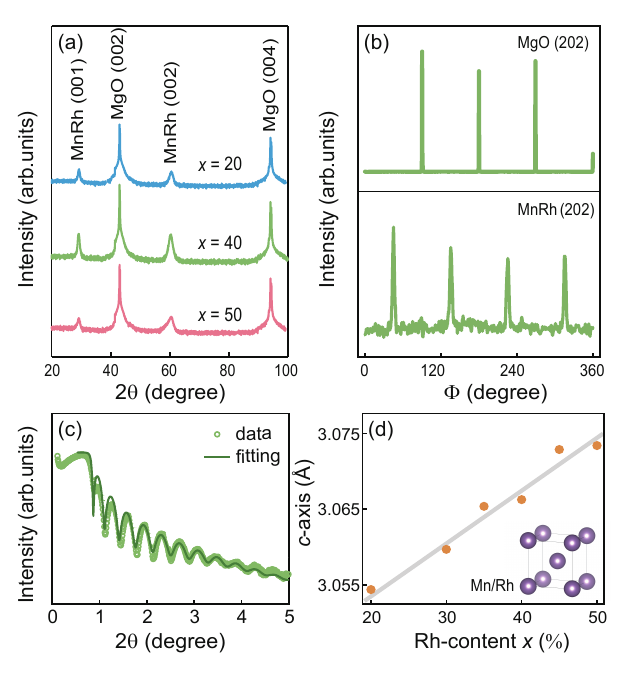}
	\caption{\label{fig:XRD}
        (a) Representative room-temperature HRXRD patterns for 
	    Mn$_{100-x}$Rh$_x$ epitaxial thin films with $x$ = 20, 40, and 50. The intensity is shown on a logarithmic scale. (b)
	    The $\mathrm{\Phi}$-scan patterns for $x$ = 40 film and MgO substrate. In-plane rotation of 45$^\circ$ for the epitaxial growth can be clearly identified. (c) The XRR pattern for $x$ = 40 film. Solid line through the data represents a fitting curve. (d) The estimated out-of-plane (i.e., $c$-axis) lattice constant of Mn$_{100-x}$Rh$_x$ films ($20 \le x \le 50$) versus Rh-content $x$. The standard deviations are within the symbols. Inset shows the cubic CsCl-type crystal structure of Mn$_{100-x}$Rh$_x$ films.}
\end{figure}
%

\section{Experimental details\label{sec:details}}\enlargethispage{8pt}
A series of Mn$_{100-x}$Rh$_x$ ($20 \le x \le 50$) thin films were epitaxially grown on (001)-oriented MgO substrates by 
magnetron co-sputtering of elemental Mn and Rh metal targets in an ultrahigh vacuum chamber with a base pressure below 5 $\times$ 10$^{-8}$ Torr.
Prior to the deposition, the MgO substrates were annealed at 600$^\circ$C for 1 hour to eliminate surface contaminants. Afterward, the substrates were cooled to 400~$^\circ$C, where both Mn and Rh atoms were deposited in an argon atmosphere with a fixed pressure of 3 mTorr.
The Mn and Rh contents were controlled by adjusting the sputtering power of Mn and Rh targets (see details in Table~\ref{tab:content}).  	
After deposition, the Mn$_{100-x}$Rh$_x$ films were annealed in situ at 650~$^\circ$C for an extra hour to improve their crystallinity. Finally, a 3-nm-thick SiN capping layer was deposited at room temperature to protect the Mn$_{100-x}$Rh$_x$ films from oxidation.

The crystal structure and the epitaxial nature of Mn$_{100-x}$Rh$_x$ films were characterized by Malvern Panalytical X’Pert high-resolution X-ray diffractometer (HRXRD). The film thickness was determined by fitting the X-ray reflectivity (XRR) patterns using the software package GenX~\cite{glavic2022}. The Mn and Rh contents were checked by energy dispersive X-ray (EDX) spectroscopy.
The magnetic properties of the Mn$_{100-x}$Rh$_x$ films were studied using a Quantum
Design magnetic properties measurement system. Measurements of transverse Hall resistivity $\rho_\mathrm{xy}$ and longitudinal resistivity $\rho_\mathrm{xx}$ were carried out in a Quantum Design physical property measurement system. For the transport measurements, all the films were patterned into a Hall-bar geometry (central area: 26~$\mu$m
$\times$ 20~$\mu$m; electrodes: 20~$\mu$m $\times$ 20~$\mu$m) by the standard photolithography and Ar-ion-beam etching techniques.
To eliminate spurious resistivity contributions due to misaligned Hall probes, the transverse contribution to the longitudinal electrical resistivity was removed by a symmetrization procedure, i.e., $\rho_\mathrm{xx}(H)$ = [$\rho_\mathrm{xx}(H)$ + $\rho_\mathrm{xx}(-H)$]/2, where $H$ is applied magnetic field. Similarly, the longitudinal contribution to the Hall resistivity was removed by an antisymmetrization procedure, $\rho_\mathrm{xy}(H)$ = [$\rho_\mathrm{xy}(H)$ -- $\rho_\mathrm{xy}(-H)$]/2.

%
\begin{figure*}[!thp]
	\centering
	\vspace{-1ex}%
	\includegraphics[width=0.85\textwidth,angle=0]{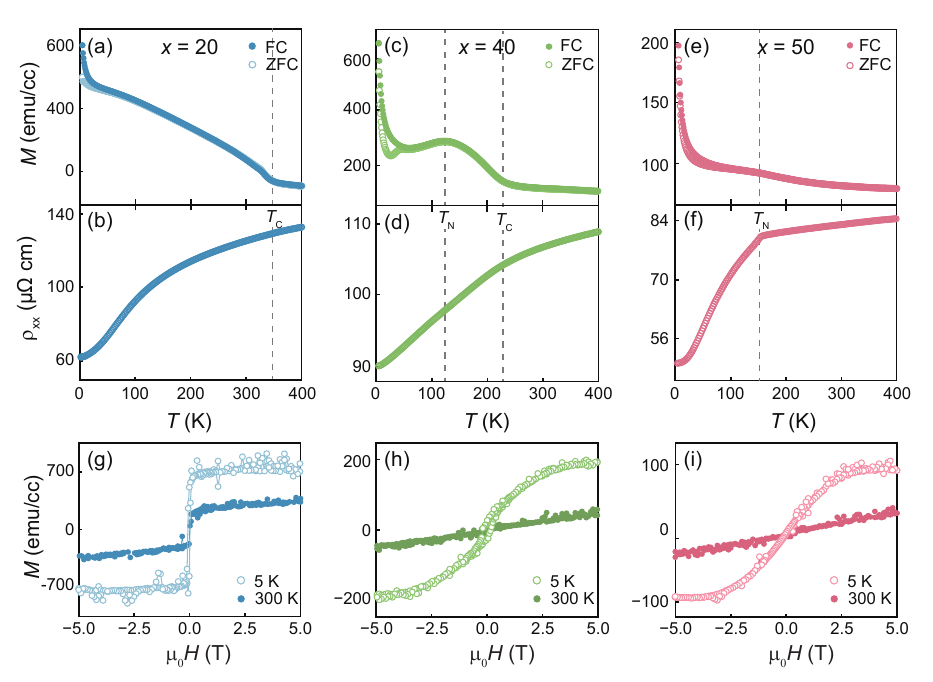}
	\caption{\label{fig:charact} \tcr{
		Temperature-dependent magnetization $M(T)$ (a) and electrical resistivity $\rho_\mathrm{xx}(T)$ (b) for Mn$_{100-x}$Rh$_x$ film with $x$ = 20. The analogous results for $x$ = 40 and 50 are shown in panels (c)-(d) and (e)-(f), respectively. 
		For $M(T)$ measurements, the magnetization was collected by applying a field of $\mu_0$$H$ = 0.1\,T within the film plane using both field-cooled (FC) and zero-field-cooled (ZFC) protocols. 
		The results for $x$ = 30, 35, and 45 are presented in Fig.~S2 in the Supplementary Materials~\cite{Supple}.
		Field-dependent magnetization $M(H)$ for $x$ = 20 (g), 40 (h), and 50 (i), respectively. The $M(H)$ data were collected at 5\,K and 300\,K by applying the magnetic field also within the film plane. \tcr{The MgO substrate contributions (Fig.~S3 in the Supplementary Materials~\cite{Supple}) were subtracted for the $M(H)$ data.}
		%
		The derivatives of electrical resistivity d$\rho_\mathrm{xx}$/d$T$ and FC-magnetization d$M$/d$T$ with respect to temperature are shown in Fig.~S4 in the Supplementary Materials~\cite{Supple}, where the magnetic transition temperatures can be clearly identified.
		}} 
\end{figure*}
%

\section{Results and discussion\label{sec:results}}\enlargethispage{8pt}
To grow the Mn$_{100-x}$Rh$_x$ films with varied Rh or Mn content, their compositions were estimated using the formula  
$x$ = $n \cdot N_\mathrm{A}$ = ($m$/$M_\mathrm{A}$) $\cdot$ $N_\mathrm{A}$ =  $[(\beta \cdot v \cdot t \cdot S)/M_\mathrm{A}] \cdot N_\mathrm{A}$,
where $n$, $m$, $N_\mathrm{A}$, and $M_\mathrm{A}$ are molar number, mass, Avogadro constant, and molar mass; $v$ and $t$ are deposition rate and time; $\beta$ and $S$ represent the density and surface area of the film, respectively. The deposition rate $v$ was controlled by adjusting the  sputtering power of Mn and Rh targets, which was calibrated by the XRR measurements. Table~\ref{tab:content} summarizes the sputtering power of Mn and Rh targets for depositing Mn$_{100-x}$Rh$_x$films with varied $x$. For instance, to produce Mn$_{80}$Rh$_{20}$ film, the sputtering power $P_\mathrm{Mn}$ and $P_\mathrm{Rh}$ were set to 50 and 15\,W, respectively. The stoichiometric ratios of all the Mn$_{100-x}$Rh$_x$ films were further characterized by EDX spectroscopy, which are consistent with the estimated Rh content in Table~\ref{tab:content}.

\begin{table}[!htb]
	\centering
	\caption{\label{tab:content}
		Summary of the sputtering power (in watt unit) of Mn ($P_{\mathrm{Mn}}$) and Rh ($P_{\mathrm{Rh}}$) targets for depositing the 
		Mn$_{100-x}$Rh$_x$films and the estimated Rh content. The deviation of Rh content is about 2\%.} 
	\begin{ruledtabular}
			\begin{tabular}{ccccccc}
					$P_{\mathrm{Mn}}$ (W) & 50 & 30 & 20 & 15 & 20 & 15 \\
					$P_{\mathrm{Rh}}$ (W) & 15 & 15 & 15 & 15 & 20 & 20 \\
					$x$ (Rh-content)      & 20 & 30 & 35 & 40 & 45 & 50\\
		\end{tabular}	
	\end{ruledtabular}
\end{table}
%

The crystal structure and the epitaxial nature of Mn$_{100-x}$Rh$_x$ films were characterized by HRXRD measurements. Figure~\ref{fig:XRD}(a) shows representative HRXRD patterns for Mn$_{100-x}$Rh$_x$ films with $x$ = 20, 40, and 50, with the other films showing similar patterns (see Fig.~S1 in the Supplementary Materials)~\cite{Supple}.
All the Mn$_{100-x}$Rh$_x$ films exhibit clear (001) and (002) reflections, confirming their cubic CsCl-type crystal structure [see inset in Fig.~\ref{fig:XRD}(d)]. The absence of foreign phases or  misorientation suggests high crystalline quality of the deposited Mn$_{100-x}$Rh$_x$ films.
The epitaxial nature of Mn$_{100-x}$Rh$_x$ films was characterized by $\mathrm{\Phi}$-scan measurements with a fixed  2$\theta$ value at the (202) reflection of MgO substrate and Mn$_{100-x}$Rh$_x$ films. The $\mathrm{\Phi}$ scans of Mn$_{100-x}$Rh$_x$ film with $x$ = 40 are plotted in Fig.~\ref{fig:XRD}(b), which confirm that the Mn$_{100-x}$Rh$_x$ films were epitaxially grown on the MgO substrate with an in-plane 45$^\circ$ rotation, i.e., Mn$_{100-x}$Rh$_x$[110](001)/MgO[100](001).
Similar $\mathrm{\Phi}$-scan patterns were observed in other Mn$_{100-x}$Rh$_x$ films (see Fig.~S1 in the Supplementary Materials)~\cite{Supple}. 
The thickness of Mn$_{100-x}$Rh$_x$ films was determined by the XRR measurements [see Fig.~\ref{fig:XRD}(c)]. The well defined finger oscillations indicate ideal flatness and uniformity of the Mn$_{100-x}$Rh$_x$ films. The determined thickness and roughness are roughly 20\,nm and 0.67\,nm for Mn$_{100-x}$Rh$_x$ films, respectively. It is noted that for $x >$ 50, the intensities of XRD reflections are significantly reduced due to the increased mismatch between Mn$_{100-x}$Rh$_x$ and MgO substrate [see Fig. S1(a) in the Supplementary Materials]~\cite{Supple}. The quality of Mn$_{100-x}$Rh$_x$ films with $x >$ 50 is less good than the films with $20 \le x \le 50$, the former might be polycrystalline in nature but with a preferred
 (00$l$) orientation. Therefore, in this work, we mainly focus on the studies of Mn$_{100-x}$Rh$_x$ films with $20 \le x \le 50$.
The out-of-plane lattice parameters (i.e., $c$-axis) of the cubic Mn$_{100-x}$Rh$_x$ ($20 \le x \le 50$) films were estimated according
to the HRXRD patterns [see Fig.~\ref{fig:XRD}(a)]. As shown in Fig.~\ref{fig:XRD}(d), the $c$-axis almost linearly increases as the Rh content $x$ increases. For $x$ =  50, the $c$ = 3.073~{\AA} is larger than the bulk value of 3.045~{\AA}~\cite{kouvel1963_17}, implying a strong strain from the MgO substrate. 

The magnetic and transport properties of Mn$_{100-x}$Rh$_x$ ($20 \le x \le 50$) films were characterized by temperature-dependent electrical resistivity $\rho_\mathrm{xx}(T)$ and magnetization $M(T)$. The representative results of $x$ = 20, 40, and 50 are presented in Fig.~\ref{fig:charact}, while the results of other films are summarized in Fig.~S2 in the Supplementary Materials~\cite{Supple}.
As indicated by the dashed lines, the magnetic transitions can be clearly identified in the $M(T)$ data. 
For $x \le 35$, the $M(T)$ resembles the FM features below $T_\mathrm{C}$ = 350\,K [Fig.~\ref{fig:charact}(a)] that is almost independent of Rh content. As $x$ further increases, $T_\mathrm{C}$ shifts to a lower temperature ($\sim$200\,K) for $x$ = 40 and 45. 
Interestingly, the $M(T)$ exhibits a peak-like anomaly at $T < T_\mathrm{C}$, implying a magnetic transition from FM to AFM state at $T_\mathrm{N}$ = 120\,K [Fig.~\ref{fig:charact}(c)]. 
Similar FM-to-AFM transition has been frequently observed in their isostructural Fe$_{100-x}$Rh$_x$ films~\cite{zhu2023_19,maat_temperature_2005,arregi_evolution_2020,mei_structural_2018}. 
Different from the Fe$_{100-x}$Rh$_x$ case, the temperature hysteresis is absent in Mn$_{100-x}$Rh$_x$ films near the magnetic transitions, confirming their second order in nature.
Finally, the FM transition disappears, and only the AFM one survives for $x$ = 50 [Fig.~\ref{fig:charact}(e)], consistent with previous results~\cite{li2025_22}. The electrical resistivity of all the Mn$_{100-x}$Rh$_x$ films decreases upon cooling the temperature, indicating their metallic feature. The magnetic transition temperatures also can be tracked in the temperature-dependent resistivity $\rho_\mathrm{xx}(T)$. For $x \le 35$, the magnetic transition at $T_\mathrm{C}$ is not so evident in the $\rho_\mathrm{xx}(T)$[Fig.~\ref{fig:charact}(b)], but still can be tracked in the derivative of resistivity with respect to the temperature d$\rho_\mathrm{xx}$/d$T$ (see Fig.~S4 in the Supplementary Materials~\cite{Supple}). For $x$ = 40 and 45, both transitions at $T_\mathrm{C}$ and $T_\mathrm{N}$ are clearly discernible in the d$\rho_\mathrm{xx}$/d$T$. As further increasing the Rh content, the $\rho_\mathrm{xx}(T)$ exhibits a distinct transition at $T_\mathrm{N}$ = 150\,K [Fig.~\ref{fig:charact}(f)]. As indicated by the dashed lines in Fig.~\ref{fig:charact}, the magnetic transition temperatures determined from $M(T)$ and $\rho_\mathrm{xx}(T)$ are highly consistent (see magnetic phase diagram below), implying their intrinsic nature.


%
\begin{figure}[b]
	\centering
	\includegraphics[width=0.48\textwidth,angle = 0]{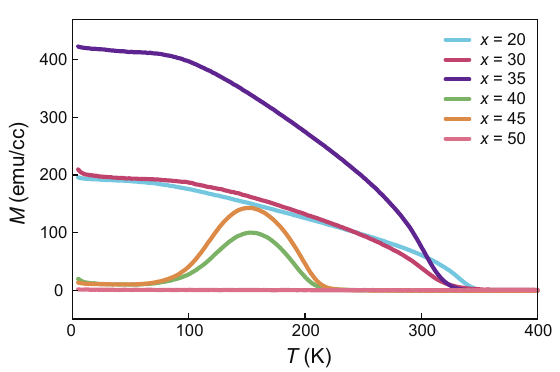}
	\caption{\label{fig:mag} \tcr{
		Temperature-dependent magnetization for Mn$_{100-x}$Rh$_x$ films ($20 \le x \le 50$), collected in the absence of external magnetic field upon heating the films. The finite magnetization indicates the presence of spontaneous magnetization in  Mn$_{100-x}$Rh$_x$ films.}}
\end{figure}
\begin{figure*}[!htp]
	\centering
	\includegraphics[width=0.85\textwidth,angle = 0]{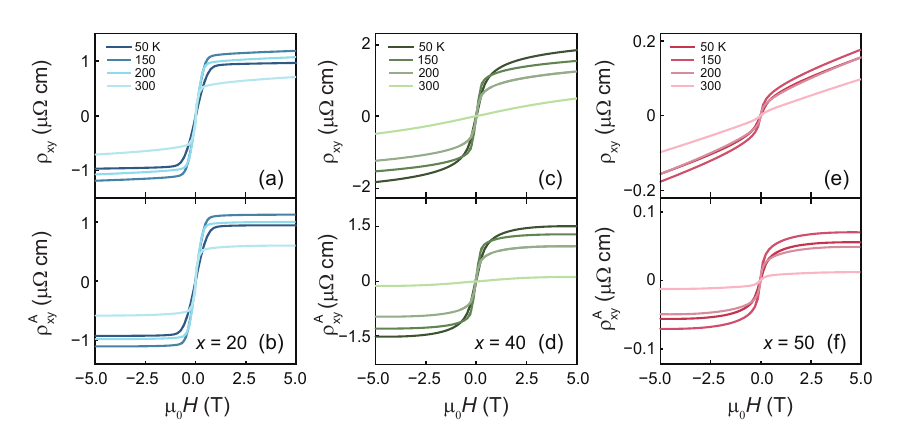}
	\caption{\label{fig:fielddep} (a) Field-dependent Hall resistivity $\rho_\mathrm{xy}(H)$ and (b) anomalous Hall resistivity $\rho^\mathrm{A}_\mathrm{xy}(H)$ at various temperatures between 50 and 300\,K for Mn$_{100-x}$Rh$_x$ films with $x$ = 20. 
		The analogous results for $x$ = 40 and 50 are shown in panels (c)-(d) and (e)-(f), respectively. 
		The $\rho^\mathrm{A}_\mathrm{xy}(H)$ at other temperatures are shown in Fig.~S5 in the Supplementary Materials~\cite{Supple}. 
		The magnetic field was applied along the normal direction of the thin film, i.e., $H \parallel c$. The results for $x$ = 30, 35, and 45 are presented in Fig. S6 in the Supplementary Materials~\cite{Supple}. }
\end{figure*}
%

The field-dependent magnetization $M(H)$ up to 5\,T was collected at 5 and 300\,K for Mn$_{100-x}$Rh$_x$ films. For $x \le 35$, \tcr{the $M(H)$ exhibits a clear square magnetic hysteresis loop. For example, the determined coercive fields of $x$ = 20 are $\mu_0$$H_c$ = 50 and 20\,mT for 5 and 300\,K, respectively, confirming the FM-type transition at $T_\mathrm{C}$ = 330-350\,K in those films.}
On the contrary, for $x > 35$, the $M(H)$ shows different field dependence. For these Mn$_{100-x}$Rh$_x$ films, 
the $M(H)$ at 300\,K is almost linear in field, consistent with their PM state at this temperature. 
When decreasing the temperature below $T_\mathrm{N}$, the $M(H)$ resembles typical features of AFM films.
As shown in Figs.~\ref{fig:charact}(h)-(i), the $M(H)$ is linear at 300\,K, consistent with the PM state, while at 5\,K it saturates at high fields, where Mn moments are gradually polarized into a field-induced FM state. 
\tcr{It is noted that the hysteresis loop observed in  Mn$_{100-x}$Rh$_x$ films with $40 \le x  \le 45$ is attributed to the remaining FM moments in the AFM state , which is consistent with the observed spontaneous magnetization in Fig.~\ref{fig:mag}. The slightly enhanced coercive field is most likely due to the pinning effect from AFM moments.} 


To further understand the nature of magnetic transitions in the Mn$_{100-x}$Rh$_x$ films, the temperature-dependent magnetization was further measured in the absence of external magnetic field. As shown in Fig.~\ref{fig:mag}, spontaneous magnetization emerges at $T_\mathrm{C}$ due to the formation of a FM order for $x \le 35$, which continuously increases and saturates as lowering the temperature.  
By contrast, for $x$ = 50, the spontaneous magnetization is almost zero throughout the entire temperature range, as expected for the AFM order. Interestingly, for $x$ = 40 and 45, similar to those with $x \le 35$ cases, the spontaneous magnetization shows up at $T < T_\mathrm{C}$. However, when the temperature decreases below $T < T_\mathrm{N}$, spontaneous magnetization starts to decrease, forming a dome-like feature in the magnetically ordered state. These $M(T)$ curves again confirm the FM-to-AFM transition in the Mn$_{100-x}$Rh$_x$ films with $x$ = 40 and 45.

Field-dependent transverse Hall $\rho_\mathrm{xy}(H)$ and longitudinal electrical resistivity $\rho_\mathrm{xx}(H)$ of Mn$_{100-x}$Rh$_x$ films were measured over a wide temperature range. The Mn$_{100-x}$Rh$_x$ films with $x$ = 20, 40, and 50 are representative examples that undergo a FM, a FM to AFM, and an AFM magnetic transition (see phase diagram below), respectively, and their $\rho_\mathrm{xy}(H)$ results are summarized in Fig.~\ref{fig:fielddep}. 
The results of $\rho_\mathrm{xx}(H)$ and $\rho_\mathrm{xy}(H)$ of other films are presented in Figs.~S5-S6 in the Supplementary Materials~\cite{Supple}. For $x \le 35$, since the $T_\mathrm{C}$ is close to or above room temperature, $\rho_\mathrm{xy}(H)$ is dominated by the AHE in the studied temperature range at $T \le$ 350\,K [see Fig.~\ref{fig:fielddep}(a) and Figs.~S5-S6 in the Supplementary Materials~\cite{Supple}. For $x$ = 40 and 45, the FM and AFM transition temperatures are close to 200 and 120\,K, respectively.
Therefore, $\rho_\mathrm{xy}(H)$ is linear in field at $T \ge$ 250\,K, where the ordinary Hall effect (OHE) is dominated.  
By contrast, $\rho_\mathrm{xy}(H)$ is again dominated by the AHE at $T <$ 250\,K for these films [see Fig.~\ref{fig:fielddep}(c)].
As the Rh content further increases, $\rho_\mathrm{xy}(H)$ of $x$ = 50 resembles the results of $x$ = 40 and 45, although the OHE is more evident in the former case [see Fig.~\ref{fig:fielddep}(e)].

\begin{figure}[!thp]
	\centering
	\includegraphics[width=0.45\textwidth,angle= 0]{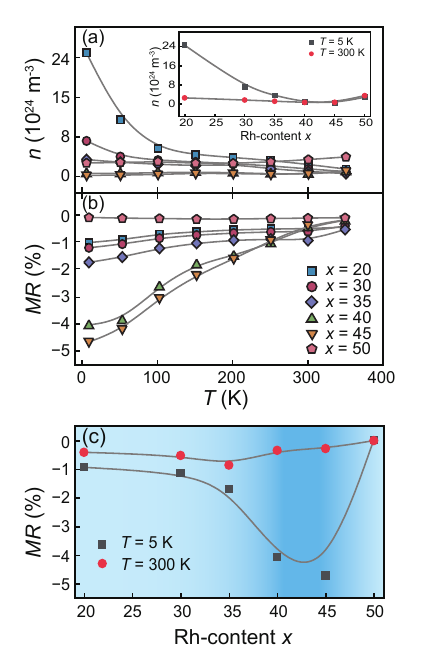}
	\caption{\label{fig:parametes} \tcr{
		Temperature-dependent carrier density $n$ (a) and magnetoresistance MR (b) for Mn$_{100-x}$Rh$_x$ films ($20 \le x \le 50$). The inset in panel (a) shows carrier density at 5\,K and 300\,K as a function of Rh content. 		
		(c) MR  versus the Rh content at $T$ =  5 and 300\,K.
		The MR was calculated according to MR = [$\rho_\mathrm{xx}$(5~T) – $\rho_\mathrm{xx}(0)$]/$\rho_\mathrm{xx}(0)$, where $\rho_\mathrm{xx}$(5~T) and $\rho_\mathrm{xx}(0)$ are the resistivity in a field of 5\,T and 0\,T, respectively.}}
\end{figure}


The Hall resistivity is fitted to $\rho_\mathrm{xy}(H)$ = $\rho_\mathrm{xy}^\mathrm{O}(H)$ +  $\rho_\mathrm{xy}^\mathrm{A}(H)$, where $\rho_\mathrm{xy}^\mathrm{O}(H)$ and $\rho_\mathrm{xy}^\mathrm{A}(H)$ are the ordinary and anomalous Hall resistivity. 
$\rho_\mathrm{xy}^\mathrm{O}$ (= $R_\mathrm{0}$$H$) is proportional to the applied magnetic field, $\rho_\mathrm{xy}^\mathrm{A}$ (= $R_\mathrm{S}$$M$) is mostly determined by the magnetization.
In real materials, $R_\mathrm{S}$ can be a constant or proportional to $\rho_\mathrm{xx}$ or $\rho_\mathrm{xx}^2$ depending on the different mechanism at play, e.g., intrinsic-, side-jump, or skew scattering~\cite{nagaosa_2010, yang_2020, chen_2021}. Considering that the magnetoresistivity (MR) is relatively small for  Mn$_{100-x}$Rh$_x$ films (Fig.~S7 in the Supplementary Materials~\cite{Supple}),   $\rho_\mathrm{xx}$ has negligible effect on $\rho_\mathrm{xy}^\mathrm{A}$. As a consequence, $\rho_\mathrm{xy}^\mathrm{A}$ was extracted simply by subtracting the linear term (i.e., $\rho_\mathrm{xy}^\mathrm{O}$).
The positive $R_0$ coefficients indicate the dominant hole-type carriers in all the Mn$_{100-x}$Rh$_x$ films.
The carrier density $n$ calculated from $R_0$ coefficient is summarized in Fig.~\ref{fig:parametes}(a). 
For $x \ge 40$, the $n$ exhibits a weak temperature dependence. For $x < 40$, the $n$ increases as the temperature decreases. 
For example, the $n$ increases from 3 $\times$ 10$^{24}$~m$^{-3}$ at 350\,K to 24 $\times$ 10$^{24}$~m$^{-3}$ at 5\,K for $x$ = 20.
Such different temperature-dependent $n(T)$ is consistent with the $\rho_\mathrm{xx}(T)$ data in Fig.~\ref{fig:charact}.
For $x$ = 20, the residual resistivity ratio (RRR) is about 2 that is almost twice larger than the rest of Mn$_{100-x}$Rh$_x$ films.
In addition, the carrier density decreases as the Rh content $x$ is increased, and exhibits a local minimum at $x$ = 40-45 [see inset in Fig.~\ref{fig:parametes}(a)]. \tcr{The $n$ is strongly correlated with the magnetic ground states of Mn$_{100-x}$Rh$_x$ films. 
In general,  Mn$_{100-x}$Rh$_x$ films with a FM ground state show a larger carrier density than the films with an AFM ground state. Therefore, the magnetic order induced band reconstruction plays the key role in determining the density of states near the Fermi level and thus the carrier density in Mn$_{100-x}$Rh$_x$ films. Further theoretical work including the band-structure calculations is highly desirable. 
}

The extracted anomalous-Hall resistivity $\rho_\mathrm{xy}^\mathrm{A}(H)$ at various temperatures are presented in the bottom panels in  Figs.~\ref{fig:fielddep}.
For $x$ = 20, the Mn$_{100-x}$Rh$_x$ film undergoes a FM transition at $T_\mathrm{C}$ = 350\,K.
As shown in Fig.~\ref{fig:fielddep}(b),
$\rho_\mathrm{xy}^\mathrm{A}(H)$ quickly saturates at $H > H_s$, which slightly increases as lowering the temperature, reaching $\sim$1\,T at 50 K. Both features are typical for FM materials. 
For $x$ = 40, it undergoes a PM-to-FM and a FM-to-AFM transition at $T_\mathrm{C}$ = 200 K  and $T_\mathrm{N}$ = 120\,K, respectively. 
Therefore, $\rho_\mathrm{xy}^\mathrm{A}$ is almost absent at room temperature. At $T_\mathrm{N}$ $<$ $T$ $<$ $T_\mathrm{C}$, $\rho_\mathrm{xy}^\mathrm{A}(H)$ behaves similarly to the FM Mn$_{100-x}$Rh$_x$ films with $x \le 35$. Finally, in the AFM state 
(i.e., $T < T_\mathrm{N}$), $\rho_\mathrm{xy}^\mathrm{A}(H)$ saturates at a relatively large field of 1.5\,T, resembling the Mn$_3$$X$ ($X$ = Sn, Ga, and Pt) noncollinear antiferromagnets~\cite{nakatsuji2015_1,liu_2017,zuniga-cespedes_2023}. As the Rh content further increases, Mn$_{100-x}$Rh$_x$ films undergo an AFM transition at $T_\mathrm{N}$ = 150\,K. 
For $x$ = 50, though the magnitude of  $\rho_\mathrm{xy}^\mathrm{A}$ is almost 30 times smaller than the other Mn$_{100-x}$Rh$_x$ films, its field dependence resembles the $\rho_\mathrm{xy}^\mathrm{A}$  of $x$ = 40 and 45 at $T < T_\mathrm{N}$.   
It is noted that the extremely weak $\rho_\mathrm{xy}^\mathrm{A}$ ($\sim$0.01~$\mu$$\mathrm{\Omega}$cm) in the PM state of  Mn$_{100-x}$Rh$_x$ film with  $x$ $>$ 35 is most likely attributed to the uncompensated magnetization at the interfaces.

\begin{figure*}[!thp]
	\centering
	\includegraphics[width=0.85\textwidth,angle= 0]{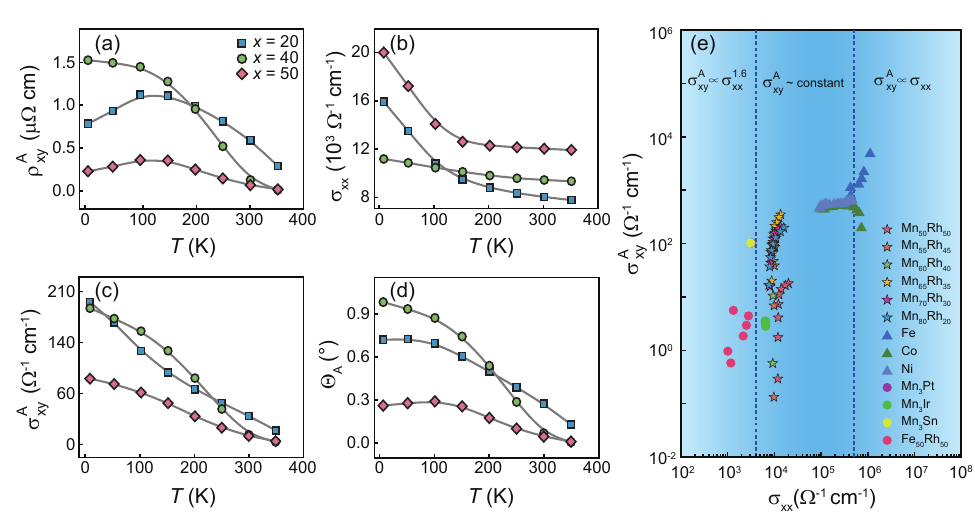}
	\caption{\label{fig:anomalous}
	 \tcr{
	     Temperature-dependent anomalous Hall resistivity $\rho_\mathrm{xy}^\mathrm{A}(T)$ (a), electrical conductivity $\sigma_\mathrm{xx}(T)$ (b), anomalous Hall conductivity $\sigma_\mathrm{xy}^\mathrm{A}(T)$ (c), and anomalous Hall angle $\Theta_\mathrm{A}(T)$ (d) for Mn$_{100-x}$Rh$_x$ films with $x$ = 20, 40, and 50. 
			$\sigma_\mathrm{xy}^\mathrm{A}$ and $\Theta_\mathrm{A}$ were calculated according to  $\sigma_\mathrm{xy}^\mathrm{A}$ = $\rho_\mathrm{xy}^\mathrm{A} /[(\rho_\mathrm{xy}^\mathrm{A})^2 + \rho_\mathrm{xx}^2]$ and $\Theta_\mathrm{A}$ = $\tan$$^{-1}$($\sigma_\mathrm{xy}^\mathrm{A}/\sigma_\mathrm{xx}$). 
			In panels (a), (c), and (d), the data for $x$ = 50 were multiplied by a factor of 5 for clarity.  
			The analogous results for x = 30, 35, and 45 are summarized in Fig.~S8 in the Supplementary Materials~\cite{Supple} 
			(e) $\sigma_\mathrm{xy}^\mathrm{A}$ vs $\sigma_\mathrm{xx}$ for various types of magnetic thin films spanning different AHE regimes, from skew-scattering mechanism ($\sigma_\mathrm{xy}^\mathrm{A}$ $\propto$ $\sigma_{xx}$), through the intrinsic- ($\sigma_\mathrm{xy}^\mathrm{A}$ $\sim$ constant), and side-jump ($\sigma_\mathrm{xy}^\mathrm{A}$ $\propto$ $\sigma_\mathrm{xx}^{1.6}$) regimes. 
			%
			Except for the Mn$_{100-x}$Rh$_x$ films (solid stars),  
			the data for other films were taken from the Refs.~\onlinecite{miyasato2007,fujishiro2021,zhu2023_19,boldrin_2019,kobayashi_2022}.
        	}}
\end{figure*}

\begin{figure}[!thp]
	\centering
	\includegraphics[width=0.45\textwidth,angle= 0]{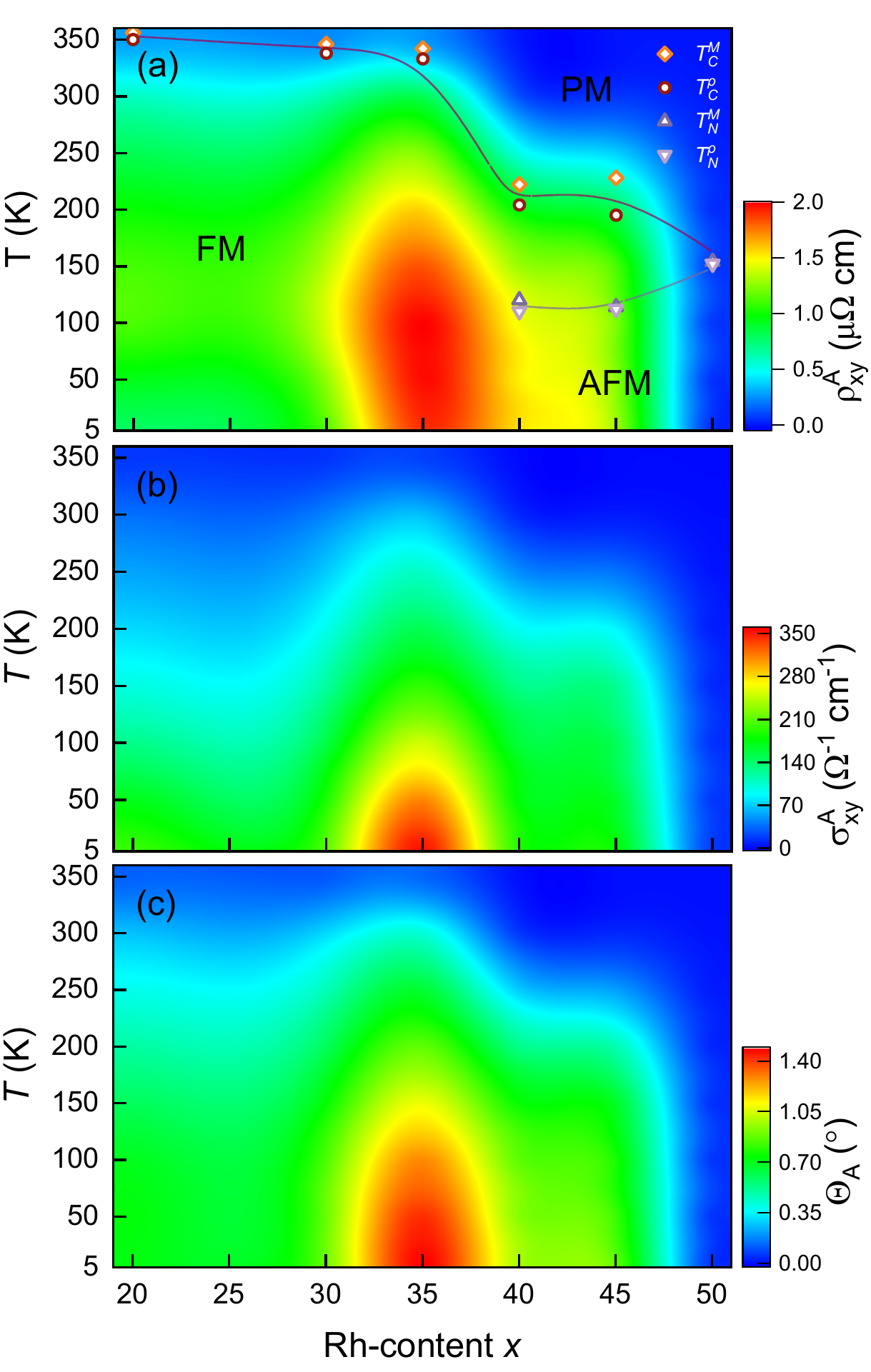}
	\caption{\label{fig:phase} 
		  \tcr{
			(a) Magnetic phase diagram of Mn$_{100-x}$Rh$_x$ films. The magnetic transition temperatures ($T_\mathrm{C}$ and $T_\mathrm{N}$) were determined from $\rho_\mathrm{xx}(T)$ and $M(T)$ (see details in Fig.~\ref{fig:charact} and Figs.~S2 and S4 in the Supplementary Materials~\cite{Supple}). The background color indicates the magnitude of extracted $\rho^\mathrm{A}_\mathrm{xy}(H)$ at various temperatures for Mn$_{100-x}$Rh$_x$ films.
			The contour plots of anomalous Hall conductivity $\sigma_\mathrm{xy}^\mathrm{A}(H,T)$ and anomalous Hall angle $\Theta_{A}(H,T)$ of  Mn$_{100-x}$Rh$_x$ films are shown in panels (b) and (c), respectively.}}
\end{figure}
	%

The temperature-dependent 5-T MR for all the Mn$_{100-x}$Rh$_x$ films is summarized in Fig.~\ref{fig:parametes}(b), while the field-dependent MR collected at various temperatures are presented in Fig.~S7 in the Supplementary Materials~\cite{Supple}.
All the Mn$_{100-x}$Rh$_x$ films exhibit negative MR in the studied temperature range. 
For $x \le 35$, the MR is weakly temperature dependent. However, for $x$ = 40 and 45, the MR is strongly temperature dependent.  
The MR ($<$0.1\%) becomes almost negligible even in the AFM state of $x$ = 50. 
The MR is small and is independent of Rh content in the PM state of Mn$_{100-x}$Rh$_x$ films [see Fig.~\ref{fig:parametes}(c)]. 
However, in the magnetically ordered state (e.g., 50\,K), the MR increases as Rh content increases, reaching a maximum value of $\sim$--5\% for $x$ = 40 and 45. Since the Mn$_{100-x}$Rh$_x$ films with $x$ = 40, 45, and 50 are antiferromagnetically ordered at temperature below $\sim$120\,K, the distinct MR values might hint at different AFM structures between $x$ = 40 (or 45) and $x$ = 50, awaiting further investigations.
The MR of  Mn$_{100-x}$Rh$_x$ films is significantly smaller than the isostructural Fe$_{100-x}$Rh$_{x}$ films~\cite{zhu2023_19,negi_formation_2023,de_vries_hall-effect_2013,suzuki_stability_2009}, which reach almost --50\% in the mixed AFM and FM states in the latter case.  
Such a giant MR of the Fe$_{100-x}$Rh$_{x}$ films is attributed to the field-induced metamagnetic transition, which is clearly absent in the Mn$_{100-x}$Rh$_x$ films.

The derived anomalous Hall resistivity $\rho_\mathrm{xy}^\mathrm{A}$ of Mn$_{100-x}$Rh$_x$ films with x = 20, 40, and 50 is summarized in Fig.~\ref{fig:anomalous}(a) as a function of temperature, while the results of other films are summarized in Fig.~S8 in the Supplementary Materials~\cite{Supple}. For $x \le$ 35, $T_C$ is between 330 and 350\,K, and thus, the $\rho_\mathrm{xy}^\mathrm{A}$ increases with decreasing temperature at $T \le$ 350\,K, and starts to saturate at temperature below 150\,K.
For $x$ = 40 and 45, a different temperature evolution was observed. The $\rho_\mathrm{xy}^\mathrm{A}$ is largely enhanced at $T \le T_\mathrm{C}$ and changes slope at temperature close to $T_\mathrm{N}$ $\sim$ 120\,K. 
For $x$ = 50, the $\rho_\mathrm{xy}^\mathrm{A}$ is significantly smaller than other Mn$_{100-x}$Rh$_x$ films. Although it shares similar features to the $x$ $\le$ 35 cases, the interfacial contribution might affect the temperature-dependent $\rho_\mathrm{xy}^\mathrm{A}$ for $x$ = 50. 
Due to the metallicity of Mn$_{100-x}$Rh$_x$ films (see Fig.~\ref{fig:charact}), both the electrical conductivity $\sigma_\mathrm{xx}$ and the anomalous Hall conductivity $\sigma_\mathrm{xy}^\mathrm{A}$ are largely enhanced at low temperatures [Fig.~\ref{fig:parametes}(b)-(c)].
For instance, $\sigma_\mathrm{xy}^\mathrm{A} \sim 360$\,$\mathrm{\Omega}^{-1}$cm$^{-1}$ at 5\,K is almost 20 times larger than
the $\sigma_\mathrm{xy}^\mathrm{A}$ $\sim$ 18\,$\mathrm{\Omega}^{-1}$cm$^{-1}$ at 350\,K for $x$ = 35. 
The anomalous Hall angle $\Theta_\mathrm{A}$ $=$ $\tan$$^{-1}$($\sigma_\mathrm{xy}^\mathrm{A}/\sigma_\mathrm{xx}$) 
is also largely enhanced at low temperatures [Fig.~\ref{fig:anomalous}(d)]. For Mn$_{100-x}$Rh$_x$ films with $x \le 45$, $\Theta_\mathrm{A}$ $\approx$ 0.8-1.5$^\circ$ at 5\,K, while it is only 0.06$^\circ$ for $x$ = 50. Such distinct $\Theta_\mathrm{A}$ values further indicate that the AFM structures are different between $x$ = 40-45 and 50. 
At room temperature, the $\Theta_\mathrm{A}$ $\approx$ 0.1$^\circ$ of Mn$_{100-x}$Rh$_x$ films ($x$ $<$ 50) is also larger than other Mn-based metallic thin films. 
For example, the isostructural Fe$_{100-x}$Rh$_x$ films with different Rh content exhibit a $\Theta_\mathrm{A}$ $<$ 0.1$^\circ$~\cite{zhu2023_19}. Despite large spin Hall conductivity has been reported in noncollinear AFM Mn$_3$Ir, its $\Theta_\mathrm{A}$ $\approx$ 0.03$^\circ$ is much smaller~\cite{kobayashi_2022}.


\tcr{The Tian-Ye-Jin (TYJ) model is widely used to distinguish the intrinsic contribution from extrinsic contributions to the AHE in magnetic materials~\cite{Tian2009}. Since there is no clear linear relationship can be identified between $\rho_\mathrm{xy}^\mathrm{A}$ and $\rho_\mathrm{xx}$ (or $\rho_\mathrm{xx}^2$) (see Fig.~S9 in the Supplementary Materials)~\cite{Supple}, the TYJ model fails to isolate the intrinsic contribution to the AHE in Mn$_{100-x}$Rh$_x$ films. The interfacial FM moments or remaining FM moments in the AFM state may also lead to temperature-dependent $\rho_\mathrm{xy}^\mathrm{A}$, which make the total $\rho_\mathrm{xy}^\mathrm{A}$ more complicated.}
As an alternative, to elucidate the mechanism of AHE in the Mn$_{100-x}$Rh$_x$ films, $\sigma_\mathrm{xy}^\mathrm{A}$ against
$\sigma_\mathrm{xx}$, together with the results of other magnetic thin films are plotted in Fig.~\ref{fig:anomalous}(e).
The scaling relation between $\sigma_\mathrm{xy}^\mathrm{A}$ and $\sigma_\mathrm{xx}$ has been frequently studied in recent years.	 
In the $\sigma_\mathrm{xy}^\mathrm{A}$ vs. $\sigma_\mathrm{xx}$ plot, three regimes with different mechanisms have been proposed to accounts for the observed $\sigma_\mathrm{xy}^\mathrm{A}$ in magnetic materials~\cite{chen_2021,yang_2020,nagaosa_2010}. First, in the high-conductivity regime ($\sigma_\mathrm{xx}$ $\gtrsim$ 5 $\times$ 10$^5$\,$\mathrm{\Omega}^{-1}$cm$^{-1}$), the extrinsic skew scattering contribution dominates AHE and $\sigma_\mathrm{xy}^\mathrm{A}$ $\propto$ $\sigma_\mathrm{xx}$. Second, in the good-metal regime (3 $\times$ 10$^3$ $\lesssim$ $\sigma_\mathrm{xx}$  $\lesssim$ 5 $\times$ 10$^5$\,$\mathrm{\Omega}^{-1}$cm$^{-1}$), $\sigma_\mathrm{xy}^\mathrm{A}$ is mostly determined by the intrinsic Berry-curvature mechanism, which is approximately independent of $\sigma_\mathrm{xx}$. Finally in the bad-metal (or localized hopping) regime ($\sigma_\mathrm{xx}$ $\lesssim$ 3 $\times$ 10$^3$\,$\mathrm{\Omega}^{-1}$cm$^{-1}$), the extrinsic side-jump mechanism is at play, which leads to $\sigma_\mathrm{xy}^\mathrm{A}$ $\propto$ $\sigma_\mathrm{xx}^{1.6-1.8}$. 
As shown by star symbols in Fig.~\ref{fig:anomalous}(e), for $T \le$ 350\,K, 
the $\sigma_\mathrm{xx}$ $\sim$ 0.7--1.7 $\times$ 10$^4$\,$\mathrm{\Omega}^{-1}$cm$^{-1}$ 
of Mn$_{100-x}$Rh$_x$ films locate at good-metal regime. Therefore, the intrinsic Berry-curvature mechanism is mostly account for the AHE in the Mn$_{100-x}$Rh$_x$ films. Similar intrinsic AHE has been found also in the noncollinear AFM Mn$_3$$X$ ($X$ = Ir,  Ge)~\cite{reis2020,nayak2016_4,kobayashi_2022}. However, the Mn$_{100-x}$Rh$_x$ films with a FM ground state (i.e., $x \le$ 35) exhibit  much larger $\sigma_\mathrm{xy}^\mathrm{A}$ than the Mn$_3$$X$. It is noted that the $\sigma_\mathrm{xx}$ of Mn$_{100-x}$Rh$_x$ films is also close to the bad-metal regime, and thus, the extrinsic side-jump contribution cannot be fully excluded. 
Besides, the weak $\sigma_\mathrm{xy}^\mathrm{A}$ caused by the uncompensated magnetization at the interfaces was observed in the PM state of Mn$_{100-x}$Rh$_x$ films [Fig.~\ref{fig:anomalous}(c)], which is rather difficult to isolate from measured $\sigma_\mathrm{xy}^\mathrm{A}$.
It could be interesting to investigate Mn$_{100-x}$Rh$_x$ films with varied thickness, where the magnetic transition temperatures, the interfacial contribution, as well as the nature of $\sigma_\mathrm{xy}^\mathrm{A}$ can be tuned.

Based on the magnetization and transport measurements, magnetic phase diagram of Mn$_{100-x}$Rh$_x$ films was built.  As shown in Fig.~\ref{fig:phase}(a), 
the phase diagram can be divided into three different regimes: I) $x < 40$, II) $40 \le x \le 45$, and III) $x > 45$. 
In regime I, Mn$_{100-x}$Rh$_x$ films undergo a FM transition below $T_\mathrm{C} \approx$ 330-350\,K; in regime II, in addition to the first FM transition at $T_\mathrm{C} \approx$ 200\,K, Mn$_{100-x}$Rh$_x$ films undergo a FM-to-AFM transition at $T_\mathrm{N} \approx$ 120\,K; finally in regime III, only one AFM transition at $T_\mathrm{N} \approx$ 150\,K can be tracked. 
In  regime I, the $T_\mathrm{C}$ decreases with increasing the Rh content $x$, while both AFM and FM transition temperatures are almost independent of $x$ in regime II [see Fig.~\ref{fig:phase}(a)]. All the $\rho_\mathrm{xy}^\mathrm{A}$, $\sigma_\mathrm{xy}^\mathrm{A}$, and $\Theta_\mathrm{A}$ of Mn$_{100-x}$Rh$_x$ films first increase as the Rh content for $x$ $\le$ 35, and starts to decrease at $x$ >$ 35$, demonstrating a dome-like feature.  
As further increasing the Rh content up to $x$ = 50, the AHE becomes much weaker.  
In general, the anomalous-Hall resistivity can be expressed as $\rho_\mathrm{xy}^\mathrm{A}$ = $R_\mathrm{S}$$M$, where $R_\mathrm{S}$ can be a constant or proportional to $\rho_\mathrm{xx}$ or $\rho_\mathrm{xx}^2$ depending on the different mechanisms at play~\cite{tian_2009}.
The magnetization at $T$ = 5\,K reaches a maximum value at $x$ = 35, and it shows almost an identical $x$ dependence as  
the $\rho_\mathrm{xy}^\mathrm{A}$ and $\sigma_\mathrm{xy}^\mathrm{A}$ (see Fig.~S10 in the Supplementary Materials~\cite{Supple}).
By contrast, the electrical resistivity $\rho_\mathrm{xx}$ at $T$ = 5\,K shows a different $x$ dependence, which exhibits a maximum value at $x$ = 40.
The $\rho_\mathrm{xx}$ is about 90~{\textmu}$\mathrm{\Omega}$cm for $x$ = 40, which slightly decreases to 50~{\textmu}$\mathrm{\Omega}$cm for $x$ = 50. \tcr{In addition, the residual resistivity $\rho_0$ exhibits a similar $x$ dependence to the resistivity at 5\,K. 
The larger $\rho_0$  in  Mn$_{100-x}$Rh$_x$  films with $40 \le x \le 45$ is most likely attributed to the extra electron magnetic scattering at the FM/AFM domain boundaries, since these films exhibit a small amount of remaining FM moments in the AFM state (see Fig.~\ref{fig:mag}). The largely reduced $\rho_0$  for $x > 45$ could be attributed to the reduced chemical disorder and electron magnetic scattering.}
Considering that the MR of Mn$_{100-x}$Rh$_x$ films is extremely small, the change in $\rho_\mathrm{xx}$ alone cannot explain the observed $x$-dependent anomalous Hall transport in Mn$_{100-x}$Rh$_x$ films. 
Since all the Mn$_{100-x}$Rh$_x$ films are located at the intrinsic regime in Fig.~\ref{fig:anomalous}(e), the $R_\mathrm{S}$ could be a constant, and is mostly determined by the Berry curvatures of their electronic bands. Theoretical calculations of the Berry curvatures and the anomalous Hall conductivity of Mn$_{100-x}$Rh$_x$ films are highly desirable to further elucidate the nature of AHE, in particular its nontrivial $x$ dependence in a wide temperature range.

\vspace{3pt}
\section{\label{ssec:Sum}Conclusion}\enlargethispage{8pt}
To summarize, we grew a series of epitaxial Mn$_{100-x}$Rh$_x$ ($20 \le x \le 50$) films on the (001)-oriented MgO substrates. 
The x-ray diffraction measurements confirm that all the Mn$_{100-x}$Rh$_x$ films exhibit a cubic CsCl-type crystal structure. 
By systematic magnetization and transport measurements, a rich magnetic phase diagram of Mn$_{100-x}$Rh$_x$ films was established.
For $x <$ 40 and $x > 45$, Mn$_{100-x}$Rh$_x$ films undergo a FM and an AFM transition, respectively. However, for $40 \le x \le 45$, Mn$_{100-x}$Rh$_x$ films undergo a subsequent FM-to-AFM transition. According to the Hall-resistivity measurements, all the Mn$_{100-x}$Rh$_x$ films exhibit distinct AHE in their magnetically ordered state. The scaling analysis between anomalous Hall conductivity and electrical conductivity hints at the intrinsic Berry-curvature mechanism for the observed AHE in Mn$_{100-x}$Rh$_x$ films. All the anomalous Hall resistivity, conductivity, and angle are closely related to the magnetization of Mn$_{100-x}$Rh$_x$ films, all reaching a maximum value at $x$ = 35. Our work suggests a strong correlation between magnetic properties and electronic band topology in Mn$_{100-x}$Rh$_x$ films. Therefore, Mn$_{100-x}$Rh$_x$ films represent one of the ideal platforms for manipulating the anomalous transport properties by Berry curvatures in the momentum space, highlighting their great potential for AFM spintronics. 

\vspace{1pt}
\begin{acknowledgments}
This work was supported by the National Natural Science Foundation of China (Grant Nos.~12374105 and 12350710785),
the Natural Science Foundation of Shanghai (Grant Nos.\ 21ZR1420500 and 21JC\-140\-2300), the Natural Science Foundation
of Chongqing (Grant No.\ CSTB-2022NSCQ-MSX1678), and the Fundamental Research Funds for the Central Universities.\\
\end{acknowledgments}



\bibliography{MnRh.bib}

\end{document}